\documentclass[aps,prd,superscriptaddress,nofootinbib,amsmath,amsfonts,preprintnumbers,notitlepage,10pt,english]{revtex4-1}
\usepackage{amsmath}
\usepackage{amssymb}
\usepackage{babel}

\makeatletter

\providecommand{\tabularnewline}{\\}

\@ifundefined{textcolor}{}
{%
 \definecolor{BLACK}{gray}{0}
 \definecolor{WHITE}{gray}{1}
 \definecolor{RED}{rgb}{1,0,0}
 \definecolor{GREEN}{rgb}{0,1,0}
 \definecolor{BLUE}{rgb}{0,0,1}
 \definecolor{CYAN}{cmyk}{1,0,0,0}
 \definecolor{MAGENTA}{cmyk}{0,1,0,0}
 \definecolor{YELLOW}{cmyk}{0,0,1,0}
 }

\@ifundefined{textcolor}{}{%
 \definecolor{BLACK}{gray}{0}
 \definecolor{WHITE}{gray}{1}
 \definecolor{RED}{rgb}{1,0,0}
 \definecolor{GREEN}{rgb}{0,1,0}
 \definecolor{BLUE}{rgb}{0,0,1}
 \definecolor{CYAN}{cmyk}{1,0,0,0}
 \definecolor{MAGENTA}{cmyk}{0,1,0,0}
 \definecolor{YELLOW}{cmyk}{0,0,1,0}
 }

\makeatother

\begin{document}

\preprint{YITP-11-17}

\title{Stability of Schwarzschild-like solutions in $f(R,\mathcal{G})$ gravity models}

\author{Antonio De Felice}
\affiliation{Department of Physics, Faculty of Science, Tokyo University of Science,
1-3, Kagurazaka, Shinjuku-ku, Tokyo 162-8601, Japan }

\author{Teruaki Suyama}
\affiliation{Research Center for the Early Universe, The University of Tokyo,
4-3-1, Hongo, Bunkyo-ku, Tokyo 113-0033, Japan}

\author{Takahiro Tanaka}
\affiliation{Yukawa Institute for Theoretical Physics, Kyoto University, Kyoto
606-8502, Japan}

\begin{abstract}
  We study linear metric perturbations around a spherically 
    symmetric static spacetime for general $f(R,\mathcal{G})$
  theories, where $R$ is the Ricci scalar and $\mathcal{G}$ is the
  Gauss-Bonnet term. We find that unless the determinant of the Hessian of
  $f(R,\mathcal{G})$ is zero, even-type perturbations have a ghost for
  any multi-pole mode. In order for these theories to be plausible
  alternatives to General Relativity, the theory should satisfy
    the condition that the ghost is massive enough to effectively
    decouple from the other fields. We study the requirement on the
    form of $f(R,\mathcal{G})$ which satisfies this condition. We also
  classify the number of propagating modes both for the odd-type and
  the even-type perturbations and derive the propagation speeds for
  each mode.
\end{abstract}

\date{\today}

\maketitle

\section{introduction}

In cosmology, modified gravity models have been considered as a possible
dynamical explanation for dark energy. These models have been constructed
in order to give late time acceleration, however it is utterly important
to see whether such modifications of gravity will also drastically change
the behavior of gravity at small scales, e.g.~in our solar system.
Among these models, some made use of the extra-dimensions \cite{Dvali:2000hr,Koyama:2005kd,Gorbunov:2005zk,Nicolis:2004qq}.
Others implemented a similar mechanism by introducing a modified kinetic-term
Lagrangian for scalar field non-minimally coupled with gravity \cite{Nicolis:2008in,Deffayet:2009wt,Lim:2010yk,Deffayet:2010qz,DeFelice:2010pv,DeFelice:2010as}.
Most of these new theories of gravity try not to introduce gravitational ghosts
in the spin-2 sector. 
Another possibility, which we consider in this paper, is to require
the Lagrangian for the gravitational sector to be a function of the
Lovelock scalars
only~\cite{Carroll:2004de,DeFelice:2006pg,Li:2007jm,DeFelice:2008wz,DeFelice:2009aj,DeFelice:2009rw,Elizalde:2010jx,DeFelice:2009ak,DeFelice:2009wp,DeFelice:2010sh,DeFelice:2010hb,DeFelice:2010hg},
\begin{equation}
S=\frac{M_{\rm Pl}^2}{2}\int d^{4}x\,\sqrt{-g}\, f(R,\mathcal{G}).
\end{equation}
 The $f(R)$ theories represent a subset of this general model \cite{Starobinsky:1980te,Capozziello:2002rd,Carroll:2003wy,Bean:2006up,Carloni:2007yv,Appleby:2007vb,Starobinsky:2007hu,Sotiriou:2008rp,DeFelice:2010aj} with an interesting phenomenology \cite{Motohashi:2009qn,Motohashi:2010sj,Motohashi:2011wn,Yagi:2009zz,Yagi:2009zm}.
This Lagrangian can be rewritten in terms of Lagrange multipliers
as
\begin{equation}
S=\frac{M_{\rm Pl}^2}{2}\int d^{4}x\sqrt{-g}\,[F\, R+\xi\,\mathcal{G}-U(F,\xi)],\label{action2}
\end{equation}
where one scalar field is coupled to $R$, the Ricci scalar, and
the other to $\mathcal{G}\equiv R^2-4R_{\alpha\beta}R^{\alpha\beta}+R_{\alpha\beta\mu\nu}R^{\alpha\beta\mu\nu}$, the Gauss-Bonnet combination.

Recently some papers appeared which shed some light on this class
of theories. In particular it was shown that on Friedmann-Lema\^\i tre-Robertson-Walker
(FLRW) backgrounds, there is a new (compared to General Relativity)
gravitational scalar mode which propagates with a scale dependent
speed of propagation. That is, its dispersion relation 
behaves like $\omega^{2}\propto k^{4}$ for large values of the wave vector $k$
\cite{DeFelice:2009ak,DeFelice:2009wp,DeFelice:2010sh,DeFelice:2010hb}.
This behavior changes the effective gravitational constant for the
matter perturbation at low redshifts. The reason for the
appearance of this $k^{4}$ term was explained in Ref.~\cite{DeFelice:2010hg}
by studying another background, the Kasner spacetime. In fact, on
this background, together with an odd mode, there are three even
modes $\psi_{i}$. The determinant of the kinetic matrix for
the even modes $A$ ($\mathcal{L}\ni A_{ij}\dot{\psi}_{i}\dot{\psi_{j}}$)
vanishes in the FLRW limit. Then, in this limit, one of the fields
can be integrated out from the Lagrangian, giving rise to a term proportional
to $k^{4}$. This result shows that the number of the degrees of freedom
for this theory depends on the background, in particular it depends
on the symmetries of the background. Furthermore, on this same background,
it was shown that, unless the Kasner manifold is very close to a
FLRW one, a propagating ghost is always present. In fact, in the
FLRW limit, it is this ghost that can be integrated out from the Lagrangian
as its mass becomes infinitely large.

In this paper we will discuss the behavior of the perturbation about
a spherically symmetric static vacuum background, whose metric can be written as
\begin{equation}
ds^{2}=g_{\mu\nu}^{0}dx^{\mu}dx^{\nu}=-A(r)\, dt^{2}+\frac{dr^{2}}{B(r)}+\frac{r^{2}\, dz^{2}}{1-z^{2}}+r^{2}(1-z^{2})\, d\varphi^{2},
\end{equation}
where $z\equiv\cos \theta$, and $\theta,\varphi$ are the standard 
  spherical coordinates.  So far, people have considered the way
  to constrain these theories by looking at the background solutions
only (approximate or not), as this step is necessary to check if local
gravity constraints are satisfied. However, it is important to
understand, and this is the goal of this study of ours, whether these
backgrounds are stable or not against linear perturbations, and what
we can learn in terms of speed of propagation and ghost issues for the
scalar gravitational modes. For such a theory, the background
equations of motion read
\begin{eqnarray}
  U & = & -{\frac{4B\,\xi'\, A'}{A\,{r}^{2}}}+{\frac{12{B}^{2}\,\xi'\, A'}{A\,{r}^{2}}}-{\frac{4B\, F'}{r}}-{\frac{2B\, F\, A'}{A\, r}}-{\frac{B\, F'\, A'}{A}}+{\frac{2F}{{r}^{2}}}-{\frac{2B\, F}{{r}^{2}}},\label{eq:Ueq}\\
  F'' & = & -{\frac{2B'\,\xi'}{{r}^{2}\, B}}-{\frac{F'\, B'}{2B}}+{\frac{B\,\xi''}{4{r}^{2}}}-{\frac{4\xi''}{{r}^{2}}}-{\frac{F\, B'}{r\, B}}+{\frac{2\xi'\, A'}{A\,{r}^{2}}}+{\frac{F'\, A'}{2A}}-{\frac{6\xi'\, B\, A'}{A\,{r}^{2}}}+{\frac{F\, A'}{A\, r}}+{\frac{6B'\,\xi'}{{r}^{2}}}\,,\\
  R & = & \frac{\partial U}{\partial F}\,,\\
  \mathcal{G} & = & \frac{\partial U}{\partial\xi}\,,
\end{eqnarray}
 where $'$ stands for differentiation with respect to $r$. We will
see that, similarly to the Kasner background, also in the present case in
general, there is one odd mode and three even modes (under the parity transformation, 
 $\theta\to \pi-\theta$ and $\varphi\to\varphi+\pi$).
As for the even modes, the kinetic matrix in
general has one negative eigenvalue, i.e.~the theory does possess
a ghost unless 1) the theory reduces to a subclass (to which $f(R)$
belongs) which satisfies the equality $f_{,RR}\, f_{,\mathcal{GG}}-f_{,R\mathcal{G}}^{2}=0$;
or 2) the background manifold is more symmetrical than
Schwarzschild, i.e.\ Minkowski or de Sitter.
These theories then face the problem of having a propagating ghost for physical backgrounds
in the scalar sector. It is however possible that for some models
introduced to explain dark energy, this mode becomes so highly massive
that it decouples from the relevant degrees of freedom at low energy.

\section{Brief review of the Regge-Wheeler-Zerilli formalism}

Before studying the metric perturbation of a spherically
symmetric static spacetime for $f(R,\mathcal{G})$ theories, let us briefly
review the formalism developed by Regge, Wheeler \cite{Regge:1957td}, and Zerilli \cite{Zerilli:1970se} to decompose
the metric perturbations according to their transformation properties
under two-dimensional rotations. 
Although Regge, Wheeler and Zerilli considered
the perturbation of the Schwarzschild spacetime (namely GR), the formalism
solely relies on the properties of spherical symmetry and can
be applied to $f(R,\mathcal{G})$ theories as well.

Let us denote the metric slightly perturbed from a spherically symmetric
static spacetime by $g_{\mu\nu}=g_{\mu\nu}^{0}+h_{\mu\nu}$. Hence $h_{\mu\nu}$ represent
infinitesimal quantities. Then, under two-dimensional
rotations on a sphere, $h_{tt},h_{tr}$ and $h_{rr}$ transform as scalars, $h_{ta}$
and $h_{ra}$ transform as vectors and $h_{ab}$ transforms as a tensor
($a,b$ are either $\theta$ or $\varphi$). Any scalar $s$  can be decomposed into the sum of spherical harmonics as
 \begin{equation}
s(t,r,\theta,\varphi)=\sum_{\ell,m}s_{\ell m}(t,r)Y_{\ell m}(\theta,\varphi),\label{scalar-decomposition}
\end{equation}
Any vector $V_{a}$ can be decomposed into a divergence part
and a divergence-free part as follows:
 \begin{equation}
V_{a}(t,r,\theta,\varphi)=\nabla_{a}\Phi_{1}+E_{a}^b\nabla_b\Phi_{2},
\end{equation}
where $\Phi_{1}$ and $\Phi_{2}$ are scalars and $E_{ab}\equiv\sqrt{\det\gamma}~\epsilon_{ab}$
with $\gamma_{ab}$ being the two-dimensional metric on the sphere
and $\epsilon_{ab}$ being the totally anti-symmetric symbol with
$\epsilon_{\theta \varphi}=1$.
Here $\nabla_{a}$ represents the covariant derivative with respect to the metric $\gamma_{ab}$.
Since $V_{a}$ is a two-component vector, it
is completely specified by the quantities $\Phi_{1}$ and $\Phi_{2}$. Then we can
apply the scalar decomposition (\ref{scalar-decomposition}) to $\Phi_{1}$
and $\Phi_{2}$ to decompose the vector quantity $V_a$ into spherical
harmonics.

Finally, any symmetric tensor $T_{ab}$ can be decomposed
as
 \begin{equation}
T_{ab}(t,r,\theta,\varphi)=\nabla_{a}\nabla_{b}\Psi_{1}+\gamma_{ab}\Psi_{2}+\frac{1}{2}\left(E_{a}{}^{c}\nabla_{c}\nabla_{b}\Psi_{3}+E_{b}{}^{c}\nabla_{c}\nabla_{a}\Psi_{3}\right),
\end{equation}
where $\Psi_{1},~\Psi_{2}$ and $\Psi_{3}$ are scalars. Since $T_{ab}$ has three
independent components, $\Psi_{1},~\Psi_{2}$ and $\Psi_{3}$ completely
specify $T_{ab}$. Then we can again apply the scalar decomposition
(\ref{scalar-decomposition}) to $\Psi_{1},~\Psi_{2}$ and $\Psi_{3}$
to decompose the tensor quantity into spherical harmonics. We 
refer to the variables accompanied by $E_{ab}$ by
 odd-type variables and the others by even-type
variables.

What makes these decompositions useful is that in the linearized equations
of motion (or equivalently, in the second order action) for $h_{\mu\nu}$,
odd-type perturbations and even-type ones completely decouple from
each other, reflecting the invariance of the background spacetime 
under parity transformation.
Therefore, we can study odd-type perturbations and even-type ones
separately as we will do in the following.

\section{Perturbation in $f(R,\mathcal{G})$ theories}

\subsection{The odd modes}

Using the Regge-Wheeler formalism, the odd-type metric perturbations
can be written as
 \begin{eqnarray}
 &  & h_{tt}=0,~~~h_{tr}=0,~~~h_{rr}=0,\\
 &  & h_{ta}=\sum_{\ell, m}h_{0,\ell m}(t,r)E_{ab}\partial^{b}Y_{\ell m}(\theta,\varphi),\\
 &  & h_{ra}=\sum_{\ell, m}h_{1,\ell m}(t,r)E_{ab}\partial^{b}Y_{\ell m}(\theta,\varphi),\\
 &  & h_{ab}=\frac{1}{2}\sum_{\ell, m}h_{2,\ell m}(t,r)\left[E_{a}^{~c}\nabla_{c}\nabla_{b}Y_{\ell m}(\theta,\varphi)+E_{b}^{~c}\nabla_{c}\nabla_{a}Y_{\ell m}(\theta,\varphi)\right].
\end{eqnarray}
Because of general covariance, not all the metric
perturbations are physical in the sense that some of them can be set
to vanish by using the gauge transformation $x^{\mu}\to x^{\mu}+\xi^{\mu}$,
where $\xi^{\mu}$ are infinitesimal. For the odd-type perturbation,
we can consider the following gauge transformation:
 \begin{equation}
\xi_{t}=\xi_{r}=0,~~~\xi_{a}=\sum_{\ell m}\Lambda_{\ell m}(t,r)E_{a}^{~b}\nabla_{b}Y_{\ell m}.
\end{equation}
By $\Lambda_{\ell m}$, we can always set $h_{2,\ell m}$ to vanish
(Regge-Wheeler gauge). By this procedure, $\Lambda_{\ell m}$ is
completely fixed and there is no remaining gauge degrees of freedom.

Then, after substituting the metric
into the action (\ref{action2}) and performing integrations by parts, 
we find that the action for the odd modes becomes 
\begin{equation}
S_{{\rm odd}}=\frac{M_{\rm Pl}^2}{2} \sum_{\ell,m}\int dt\, dr\,{\cal L}_{{\rm odd}}=\frac{M_{\rm Pl}^2}{2} \sum_{\ell,m}\int dt\, dr\,\bigg[A_{1}{\left({\dot{h}_{1}}-h_{0}'\right)}^{2}+A_{2}h_{0}{\dot{h}_{1}}+A_{3}h_{0}^{2}-A_{4}h_{1}^{2}\bigg],\label{odd-action}
\end{equation}
omitting the suffixes $\ell$ and $m$ for the fields, and 
\begin{eqnarray}
A_{1} & = & j^{2}\,\frac{(r\, F-4B\,\xi')}{2r}\,\sqrt{\frac{B}{A}}\,,\\
A_{2} & = & \frac{4A_{1}}{r}\,,\\
A_{3} & = & \frac{1}{r^{2}}\left[2rA_{1}'+2A_{1}+\frac{j^{2}(j^{2}-2)}{2\sqrt{AB}}(F-2B'\,\xi'-4B\,\xi'')\right],\\
A_{4} & = & \frac{j^{2}}{2}\,\frac{(j^{2}-2)\,(A\, F-2B\, A'\,\xi')}{r^{2}}\sqrt{\frac{B}{A}}\,,
\end{eqnarray}
where $j^{2}=\ell\,(\ell+1)$. Since no time derivative of $h_{0}$
appears, variating with respect to $h_{0}$ yields a constraint equation. However,
because of the presence of $h_{0}'$ in the action, the constraint
results into a second order ordinary differential equation for $h_{0}$:
 \begin{equation}
[A_{1}(h_{0}'-{\dot{h}_{1}})]'=A_{3}\, h_{0}+\frac{1}{2}\, A_{2}\,{\dot{h}_{1}}\,,\label{cons}
\end{equation}
which cannot be immediately solved for $h_{0}$. Hence, we take the
following steps to overcome this obstacle.

Let us first rewrite the action as 
\begin{equation}
\mathcal{L}_{{\rm odd}}=A_{{1}}\left(\dot{h}_{{1}}-h'_{{0}}+2\,{\frac{h_{{0}}}{r}}\right)^{2}-\frac{2(A_{1}+rA_{1}')}{r^{2}}h_{0}^{2}+A_{{3}}{h_{{0}}}^{2}-A_{{4}}{h_{{1}}}^{2}.\label{eq:Lodd2}
\end{equation}
so that all the terms containing $\dot{h}_{1}$ are inside the first squared term. 
Using a Lagrange multiplier $Q$, we rewrite Eq.~(\ref{eq:Lodd2})
as follows
\begin{equation}
\mathcal{L}_{{\rm odd}}=A_{{1}}\left[2\, Q\left(\dot{h}_{{1}}-h'_{{0}}+2\,{\frac{h_{{0}}}{r}}\right)-Q^{2}\right]-{\frac{2\left(A'_{{1}}r+A_{{1}}\right){h_{{0}}}^{2}}{{r}^{2}}}+A_{{3}}{h_{{0}}}^{2}-A_{{4}}{h_{{1}}}^{2}\,.\label{eq:Lodd3}
\end{equation}
Then, both fields $h_{0}$ and $h_{1}$ can be integrated out by using
their own equations of motion, which can be written as
\begin{eqnarray}
h_{1} & = & -\frac{A_{1}\,\dot{Q}}{A_{4}}\,,\label{eq:oddh1}\\
h_{0} & = & \frac{r}{2A_{1}+2r\, A'_{1}-A_{3}r^{2}}\,[(r\, A'_{1}+2A_{1})\, Q+r\, A_{1}\, Q']\:.\label{eq:oddh0}
\end{eqnarray}
 These relations link the physical modes $h_{0}$ and $h_{1}$ to
the auxiliary field $Q$. Once $Q$ is known also $h_{0}$ and $h_{1}$
are. After substituting these expressions into the Lagrangian and performing
an integration by parts for the term proportional to $Q'\, Q$, one finds the Lagrangian
in the canonical form
\begin{equation}
\mathcal{L}_{{\rm odd}}=\frac{A_{1}^{2}}{A_{4}}\,\dot{Q}^{2}-\frac{A_{1}^{2}\, r^{2}}{A_{3}\, r^{2}-2r\, A'_{1}-2A_{1}}\,(Q')^{2}-\mu^{2}\, Q^{2}\,,\label{eq:LoddF}
\end{equation}
 where
\begin{equation}
\mu^{2}={\frac{A_{{1}}{r}^{2}\left({r}^{2}A'_{{1}}A'_{{3}}-{r}^{2}A''_{{1}}A_{{3}}+2\, A_{{1}}A_{{3}}+4\,{A'_{{1}}}^{2}+{A_{{3}}}^{2}{r}^{2}-2\, A_{{1}}A''_{{1}}+2\, A_{{1}}rA'_{{3}}-4\, A'_{{1}}rA_{{3}}\right)}{\left(2\, A_{{1}}+2\, A'_{{1}}r-A_{{3}}{r}^{2}\right)^{2}}}\,.
\end{equation}
From Eq.~(\ref{eq:LoddF}), we can derive the no ghost condition
\[
A_{4}\geq0\,,\qquad{\rm or \, \, equivalently\qquad}A\, F-2\, B\xi'\, A'\geq0\,.
\]
For solutions proportional to $e^{i(\omega t-kr)}$ with large
$k$ and $\omega$, we have the radial dispersion relation
\[
\omega^{2}=\frac{B\,\left(A\, F-2\, B\xi'\, A'\right)}{(F-4B\xi''-2B'\xi')}\, k^{2}\,,
\]
 where we made use of the background equations of motion. Finally
the expression for the radial speed reads

\[
c_{{\rm odd}}^{2}=\left(\frac{dr_{*}}{d\tau}\right)^2=\frac{\left(A\, F-2\, B\xi'\, A'\right)}{A\,(F-4B\xi''-2B'\xi')}\,,
\]
 where we used the radial tortoise coordinate ($dr_{*}^{2}=dr^{2}/B$)
and the proper time ($d\tau^{2}=A\, dt^{2}$). Therefore in order
for the modes to be stable, one also requires
\[
F-4B\xi''-2B'\xi'\geq0\,.
\]

\subsection{Even modes}

Now that we have got an idea how the action approach works for the
odd modes we can tackle the more complicated problem of the even modes.
In this case, the perturbed metric can be written as
 \begin{eqnarray}
 &  & h_{tt}=-A(r)\sum_{\ell, m}H_{0,\ell m}(t,r)Y_{\ell m}(\theta,\varphi),\\
 &  & h_{tr}=\sum_{\ell, m}H_{1,\ell m}(t,r)Y_{\ell m}(\theta,\varphi),\\
 &  & h_{rr}=\frac{1}{B(r)}\sum_{\ell, m}H_{2,\ell m}(t,r)Y_{\ell m}(\theta,\varphi),\\
 &  & h_{ra}=\sum_{\ell, m}\alpha_{\ell m}(t,r)\partial_{a}Y_{\ell m}(\theta,\varphi)\,,
\end{eqnarray}
and we use the gauge transformation to set $h_{ta}$ and $h_{ab}$ to vanish. 
In addition to the metric perturbations,
we need to perturb also the extra scalar fields $F$ and $\xi$ as
\begin{equation}
F=F(r)+\sum_{\ell, m}\delta F_{\ell m}(t,r)Y_{\ell m}\,,\qquad\mathrm{and}\qquad\xi=\xi(r)+\sum_{\ell, m}\delta\xi_{\ell m}(t,r)Y_{\ell m}\,.
\end{equation}
Then, the action at second order for the even modes, reads as follows
\begin{equation}
S_{\rm even}=\frac{M_{\rm Pl}^2}{2} \sum_{\ell,m}\int dt\, dr\,\mathcal{L}_{\rm even}\,,
\end{equation}
 where
\begin{eqnarray}
\mathcal{L}_{\rm even} & = & H'_{0}\,(a_{1}\delta\xi'+a_{2}\,\delta F'+a_{3}H_{2}+j^{2}\, a_{4}\,\alpha+a_{5}\,\delta\xi+a_{6}\,\delta F)+j^{2}\, H_{0}\,(a_{7}\, H_{2}+a_{8}\,\alpha+a_{9}\,\delta\xi+a_{10}\,\delta F)\nonumber \\
 &  &+ j^{2}\, b_{1}\, H_{1}^{2}+H_{1}\,(b_{2}\,\dot{\delta\xi}'+b_{3}\,\dot{\delta F}'+b_{4}\,\dot{H}_{2}+j^{2}\, b_{5}\,\dot{\alpha}+b_{6}\dot{\delta\xi}+b_{7}\dot{\delta F})+c_{1}\, H_{2}^{2}\nonumber \\
 & &+  H_{2}\,[c_{2}\,\delta\xi'+c_{3}\,\delta F'+j^{2}\, c_{4}\,\alpha+\delta\xi\,(j^{2}\, c_{6}+c_{7})+\delta F\,(j^{2}\, c_{9}+c_{10})]+\dot{H}_{2}\,(c_{5}\,\dot{\delta\xi}+c_{8}\,\dot{\delta F})\nonumber \\
 &  &+ j^{2}\,(d_{1}\,\dot{\alpha}^{2}+d_{2}\,\alpha^{2})+j^{2}\,\alpha\,(d_{3}\,\delta\xi'+d_{4}\,\delta F'+d_{5}\,\delta\xi+d_{6}\,\delta F)+e_{1}\,\delta F^{2}+e_{2}\,\delta F\,\delta\xi+f_{1}\,\delta\xi^{2}\,,\label{eq:azioEv}
\end{eqnarray}
where $a_{i},b_{i},c_{i},d_{i},e_{i}$
and $f_{1}$ are all functions of $r$ only and their expressions
are given in the Appendix. For simplicity, we omitted the subscripts $\ell,m$ also for the even modes. 
In what follows, we will integrate out the fields $H_{0}$, $H_{1}$, and $H_{2}$.

We first integrate out the non-propagating field $H_{1}$, by using
its own equation of motion
 \begin{equation}
H_{1}=-\frac{1}{2j^{2}b_{1}}\,(b_{2}\,\dot{\delta\xi}'+b_{3}\,\dot{\delta F}'+b_{4}\,\dot{H}_{2}+j^{2}\, b_{5}\,\dot{\alpha}+b_{6}\,\dot{\delta\xi}+b_{7}\,\dot{\delta F})\,.
\end{equation}
We note that the term proportional to $H_{0}^{2}$ is absent 
in the action.
Thus, the equation of motion for $H_0$ 
\begin{eqnarray}
a_{1}\,\delta\xi'' & + & a_{2}\,\delta F''+(a_{5}+a'_{1})\,\delta\xi+(a'_{2}+a_{6})\,\delta F+j^{2}\, a_{4}\,\alpha'+a_{3}\, H'_{2}+(a'_{5}-j^{2}\, a_{9})\,\delta\xi+(a'_{6}-j^{2}\, a_{10})\,\delta\xi\nonumber \\
 & + & (a'_{3}-j^{2}\, a_{7})\, H_{2}+(a'_{4}-a_{8})\, j^{2}\,\alpha=0\,,\label{eq:cnstH0}
\end{eqnarray}
sets a constraint for the other fields.
By looking at this equation, one may think that we cannot use it to 
directly substitute back any of the fields. However, we can remove the highest $r$-derivatives
for the fields $\delta F$, $\delta\xi$, $\alpha$ and $H_{2}$, by performing
the following field redefinition
\begin{eqnarray}
j^{2}\, a_{4}\,\alpha & = & a_{4}\, v_{0}-a_{3}\, H_{2}-a_{2}\, v'_{1}\,.\label{eq:alv0}\\
\delta F & = & v_{1}-\frac{a_{1}}{a_{2}}\,\delta\xi=v_{1}-\frac{4(1-B)}{r^2}\,\delta\xi\,.
\end{eqnarray}
Because of this field redefinition, Eq.~(\ref{eq:cnstH0})
has no more second $r$-derivative for any of the fields, and
no $r$-derivatives for the field $H_{2}$. Solving for $H_2$, we obtain
\begin{eqnarray}
\left(j^{2}\, a_{7}-\frac{a_{8}\, a_{3}}{a_{4}}\right)H_{2} & = & \left(a_{6}+\frac{a_{2}\, a_{8}}{a_{4}}\right)\, v'_{1}+\left(a_{5}-a'_{1}+\frac{a'_{2}\, a_{1}}{a_{2}}-\frac{a_{1}\, a_{6}}{a_{2}}\right)\,\delta\xi'+a_{4}\, v'_{0}+(a'_{6}-j^{2}\, a_{10})\, v_{1}+(a'_{4}-a_{8})\, v_{0}\nonumber \\
 & + & \left[a'_{5}-j^{2}\, a_{9}+\frac{a'_{1}\,(a'_{2}-a_{6})}{a_{2}}+\frac{a_{1}\,(a''_{2}-a'_{6})}{a_{2}}+\frac{a_{1}\, a'_{2}\,(a_{6}-a'_{2})}{a_{2}^{2}}-a''_{1}+\frac{j^{2}\, a_{1}\, a_{10}}{a_{2}}\right]\delta\xi\,.
\end{eqnarray}
 We can now substitute $H_{2}$ back into the action, so that also
the field $H_{0}$ is automatically eliminated. We will find it convenient
to finally perform the field substitution
\begin{eqnarray}
v_{0} & = & v_{2}\,(1+4j^{2})^{1/2},\label{eq:v0JJ}\\
\delta\xi & = & v_{3}\,(1+4j^{2})^{1/2},\label{eq:fieldredef3}
\end{eqnarray}
where the dependence on $j$ in Eqs.~(\ref{eq:v0JJ}) to (\ref{eq:fieldredef3})
is chosen such that, for large $j$, the ghost conditions, which will
be found below, become independent of $j$. Now the Lagrangian takes
the canonical form
\begin{equation}
\mathcal{L}_{\rm even}=\sum_{i,j=1}^{3}[K_{ij}(r,j)\,\dot{v}_{i}\,\dot{v}_{j}-L_{ij}(r,j)\, v'_{i}\, v'_{j}-D_{ij}(r,j)\, v'_{i}\, v_{j}-M_{ij}(r,j)\, v_{i}\, v_{j}]\,,
\end{equation}
 where $i,j$ run from 1 to 3, and the coefficients important for our
discussion are given in appendix. All matrices are symmetric except
for $D_{ij}$, which is anti-symmetric. Now we can discuss the existence
of ghosts and the speed of propagation for the modes. The no-ghost condition requires
the matrix $\mathbf{K}$ to be positive definite, that is
\begin{equation}
K_{33}>0\,,\qquad K_{22}\, K_{33}-K_{23}^{2}>0\,,\qquad\det(K_{ij})>0.
\end{equation}
 We find that $K_{22}\, K_{33}-K_{23}^{2}$ is given by
 \begin{equation}
K_{22}K_{33}-K_{23}^{2}=-\frac{16\,(1+4j^{2})^{2}\, A\, B\,(2B-2-r\, B')^{2}\,[4(B-1)\,\xi'-r^{2}F']^{2}}{r^{2}\,\Delta^{2}}\leqq0\,,\label{eq:evenghost}
\end{equation}
 where
\begin{equation}
\Delta=24AB^{2}\xi'-12B^{2}rA'\xi'-8AB\xi'+4r\xi'BA'-4ABFr+2BFA'r^{2}-2ABF'r^{2}+BA'r^{3}F'+j^{2}(2AFr-8AB\xi')\,.\label{eq:Delta}
\end{equation}
Therefore, on this background, a ghost is always
present. The determinant of the kinetic matrix is given by
\begin{equation}
\det(K_{ij})=-32\sqrt{AB}\,\frac{(j^{2}-2)(1+4j^{2})^{2}r^{2}(2B-2-rB')^{2}\,(F-2B'\xi'-4B\xi'')}{j^{2}\,\Delta^{2}}\,.\label{eq:detEv}
\end{equation}
This quantity gives us new information. Indeed, on backgrounds of
exact solutions given by $B=1+C\, r^{2}$ with $C$ being constant
(Minkowski or de Sitter solution), both the determinant and the ghost
kinetic term vanish, which implies an effective reduction of the
degrees of freedom similar to what occurs on FLRW background, where
the missing degree of freedom is ghost-like
\cite{DeFelice:2010hg}. The results obtained for the odd modes impose
$F-2B'\xi'-4B\xi''\geq0$.  Then, for the even-modes,
$\det(K_{ij})<0$ holds, so that either there is one ghost ($K_{33}>0$)
or there are three ghosts ($K_{33}<0$) on the backgrounds of general
spherically symmetric static solutions in these modified gravity
theories. The cases with $\ell=0$ and $\ell=1$ should be treated with
care, as some variables are absent from the beginning due to the
  lack of vector (for $\ell=0$) and tensor (for $\ell\leq 1$)
harmonics: for example, there is no contribution from $\alpha$ in the
action for $j=0$.

Assuming the solutions in the form proportional to $e^{i\omega t-ikr}$,  
the dispersion relation for large $k$ and $\omega$
can be obtained by solving the discriminant
\begin{equation}
\det(\omega^{2}\, A_{ij}-k^{2}\, D_{ij})=0\, ,
\end{equation}
which is a cubic equation in $\omega^{2}$.
The three radial speeds of propagation we obtain are
\begin{eqnarray}
c_{1}^{2} & = & c_{2}^{2}=\frac{\left(2A\, B-2\, A-r\, B\, A'\right)}{\left(2\, B-2-r\, B'\right)A}\,,\label{eq:speed1}\\
c_{3}^{2} & = & \frac{A\, F-2\, A'\, B\,\xi'}{A\,(F-2B'\,\xi'-4B\,\xi'')}\,.
\end{eqnarray}
 Two of the speeds of propagation reduce to unity for 
 backgrounds with $A=B$.
The third one, which is identical to the one for the odd modes, depends
directly on the profile of the two new scalar degrees of freedom
 $F$ and $\xi$ even if we set $A=B$.
The speed of propagation for large $j$, does not have a simple analytical
form. Although one can solve a cubic equation in  $\omega^{2}$, the expression
is too complicated to gain intuition from it.

\subsection{Discussion regarding the ghost}

We have found that at least one ghost mode is always present around
the spherically symmetric static background in vacuum. However, the existence of the
ghost does not necessarily mean that the background spacetime is unstable
due to the creation of the ghost and normal particle pairs. Our implicit
assumptions are that there is an yet unknown complete theory (maybe
string theory), which is well-defined at any energy scale and does not
have any ghost and that the $f(R,\mathcal{G})$ theories are the
derived effective theories, which are valid only below some cutoff
scale $M_{{\rm cutoff}}$. What is generally thought is that if the
mass of the ghost is always heavier than $M_{{\rm cutoff}}$, such
a ghost should not be regarded as a physical mode and must be integrated
out to have more sensible effective theories. On the other hand, if
the mass of the ghost becomes lighter than $M_{{\rm cutoff}}$ in
certain situations, such theories do not make sense and must be ruled
out from the list of the possible low energy effective gravitational theories.

In the models we study here, there are in general two ways out of
which the ghost mode can become massive. One is due to symmetry. That
is, as the background becomes more and more similar to either Minkowski,
de Sitter or FLRW, its mass tends to infinity because its kinetic
term vanishes. The other is due to the so-called chameleon mechanism.
Thanks to the local value of $R$ or $\mathcal{G}$ much larger than
the corresponding cosmological value, some modes may develop their
masses. Our cutoff mass $M_{{\rm cutoff}}$ is not necessarily as
large as Planck scale, but $M_{{\rm cutoff}}^{-1}$ must be sufficiently
smaller than the experimentally relevant length scale $L_{{\rm exp}}$.

Let us give approximate values for the masses of the modes, including
the ghost mode. Assuming the background around a star to be very close
to the standard GR case, then one has $F\approx1$, or $F'\approx0$,
and $\xi'\approx0\approx\xi''$. Further we assume that the theories
satisfy solar system constraints, that is, $A\approx B\approx1-r_{s}/r$,
(where $r_{s}$ is the Schwarzschild radius of the star). Under these
assumptions, the leading contribution in the mass matrix is given
by the terms originating from $U_{,FF}$, $U_{,F\xi}$, and $U_{,\xi\xi}$,
\[
M_{11}=\frac{1}{2}\, U_{,FF}\, r^{2}\,,\quad M_{13}=-\frac{(1+4j^{2})^{1/2}}{2}\,\frac{4r_{s}\, U_{,FF}-r^{3}\, U_{,F\xi}}{r}\,,\quad M_{33}=\frac{1+4j^{2}}{2r^{4}}\,(U_{,\xi\xi}r^{6}-8U_{,F\xi}r_{s}r^{3}+16r_{s}^{2}\, U_{,FF})\,.
\]
 Then the discriminant equation, for low $k$, to solve is
 \begin{equation}
\det(m^{2}\, A(r)\, K_{ij}-M_{ij})=0\,,\label{eq:discr1}
\end{equation}
 where the factor $A(r)$ comes because of the choice of the proper
time as time variable, and the elements $K_{ij}$ are given in Appendix.
Equation~(\ref{eq:discr1}) reduces to 
\begin{equation}
m^{2}\,\left[9\tilde{{\cal G}}\, m^{4}+3(U_{,\xi\xi}-\tilde{{\cal G}}U_{,FF})\, m^{2}-(U_{,FF}\, U_{,\xi\xi}-U_{,F\xi}^{2})\right]=0\,,\label{eq:discrM}
\end{equation}
where we introduced $\tilde{{\cal G}}\equiv 4{\cal G}/3\approx16r_{s}^{2}/r^{6}$.
It is also suggestive to rewrite the above equation in terms of the
function $f(R,{\cal G})$, by using the relation
 \begin{equation}
\left(\begin{array}{cc}
U_{,FF} & U_{,F\xi}\\
U_{,F\xi} & U_{,\xi\xi}
\end{array}\right)=\left(\begin{array}{cc}
f_{,RR} & f_{,R{\cal G}}\\
f_{,R{\cal G}} & f_{,{\cal GG}}
\end{array}\right)^{-1}~,
\end{equation}
 as
 \begin{equation}
m^{2}\,\left[9\tilde{{\cal G}}\det(f_{,ij})\, m^{4}+3(f_{,RR}-\tilde{{\cal G}}f_{,{\cal GG}})\, m^{2}-1\right]=0\,.\label{eq:discrM2}
\end{equation}

One obvious solution of Eq.~(\ref{eq:discrM}) is $m^{2}=0$, whose eigenvector
is given by $\mathbf{E}_{0}\equiv[0,1,0]^{t}$. The kinetic term of
this mode
 \[
\mathbf{E}_{0}^{t}\mathbf{A}\,\mathbf{E}_{0}=\frac{2(r-r_{s})^{2}\,(1+4j^{2})\,(j^{2}-2)}{j^{2}\,[3r_{s}+(j^{2}-2)\, r]^{2}}\,,
\]
is positive for $\ell\geq2$. Thus, this mode represents the GR standard
contribution. The other two mass eigenvalues are given by 
\begin{equation}
m_{\pm}^{2}=\frac{-U_{,\xi\xi}+\tilde{{\cal G}}U_{,FF}\pm(U_{,\xi\xi}+\tilde{{\cal G}}U_{,FF})\sqrt{1-\frac{4\tilde{{\cal G}}U_{,F\xi}^{2}}{(U_{,\xi\xi}+\tilde{{\cal G}}U_{,FF})^{2}}}}{6\tilde{{\cal G}}}~.\label{msquare}
\end{equation}
In order to study which of these mass eigenvalues correspond to the
ghost mode, let us perform a little more detailed analysis. First,
we diagonalize the kinetic matrix $A\, K_{ij}$ by arranging linear
combinations of fields $v_{i}=P_{ij}\, w_{j}$, which are explicitly
expressed as
 \begin{eqnarray}
v_{1} & = & \frac{\sqrt{6}}{3r}\, w_{1}+\frac{2}{r\,\sqrt{6}}\, w_{2}\,,\\
v_{2} & = & \frac{r\, j^{2}+r_{s}}{\sqrt{6}\,\sqrt{1+4j^{2}}\,(r-r_{s})}\, w_{1}+\frac{r\, j^{2}+7r_{s}-6r}{\sqrt{6}\,\sqrt{1+4j^{2}}\,(r-r_{s})}\, w_{2}+w_{3}\,,\\
v_{3} & = & \frac{r^{2}}{2\sqrt{6}\,\sqrt{1+4j^{2}}\, r_{s}}\, w_{2}\,.
\end{eqnarray}
Then, the new diagonalized kinetic matrix $\tilde{K}_{ij}\equiv A(r)\, P_{ki}\, K_{kl}\, P_{lj}$
takes its diagonal elements
 \[
\tilde{K}_{11}=1\,,\quad\tilde{K}_{22}=-1\,,\quad\tilde{K}_{33}=\frac{2(r-r_{s})^{2}\,(1+4j^{2})\,(j^{2}-2)}{j^{2}\,[(j^{2}-2)\, r+3r_{s}]^{2}}\,,
\]
whereas the new symmetric mass matrix, $\tilde{M}_{ij}=P_{ki}\, M_{kl}\, P_{lj}$,
satisfies $\tilde{M}_{i3}=0$ for $i=1,2,3$, and its non-zero components
are
 \[
\tilde{M}_{11}=\frac{U_{,FF}}{3}\,,\quad\tilde{M}_{12}=\frac{U_{,F\xi}}{3\sqrt{\tilde{{\cal G}}}}\,,\quad\tilde{M}_{22}=\frac{U_{,\xi\xi}}{3\tilde{{\cal G}}}\,.
\]
 At this point, one can easily see that $\tilde{M}_{22}$ is the most
enhanced mass-element in the Minkowski limit, $r_{s}\to0$, which gives a divergent
mass to the ghost mode. This is consistent with the notion that the
ghost possesses a divergent mass in the Minkowski background. However,
one possible natural hierarchy among the components of $\tilde{M}_{ij}$
will be that $\tilde{M}_{11}$, $\tilde{M}_{12}$ and $\tilde{M}_{22}$
are the same order when $\tilde{{\cal G}}=O(H_{0}^{4})$, where $H_{0}$
is the present value of the Hubble parameter. In this case, since
the value of $\tilde{{\cal G}}$ around the local gravitational source
is typically much larger than the cosmological value $H_{0}^{4}$,
the simple limit $\tilde{{\cal G}}\to0$, which relatively enhances
$\tilde{M}_{22}$, is out of the relevant parameter range.

Since the GR mode $w_{3}$ completely decouples from the others, we
shall concentrate on $w_{1}$ and $w_{2}$ below. They are still coupled
through the off-diagonal mass matrix ${\tilde{M}}_{12}$. The field
transformation which keeps the kinetic matrix unchanged is given by
$w_{i}=Z_{ij}z_{j}$, where $Z_{11}=Z_{22}=\cosh\beta$, and $Z_{12}=Z_{21}=\sinh\beta$.
Then, the condition for the off-diagonal component of the new mass
matrix to vanish yields
 \begin{equation}
\tanh2\beta=-\frac{2\sqrt{\tilde{{\cal G}}}U_{,F\xi}}{U_{,\xi\xi}+\tilde{{\cal G}}U_{,FF}}.\label{eq-beta}
\end{equation}
Therefore, $4\tilde{{\cal G}}U_{,F\xi}^{2}<(U_{,\xi\xi}+\tilde{{\cal G}}U_{,FF})^{2}$
is required to have the mass matrix diagonalized. In terms of $f$,
this condition is equivalent to $-4\tilde{{\cal G}}\det(f_{,ij})<(f_{,RR}-\tilde{{\cal G}}f_{,{\cal GG}})^{2}$.
In this case, the new fields can be identified as independent decoupled
massive modes, and 11 and 22 components of the diagonal mass matrix
are identified with $m_{+}^{2}$ and $-m_{-}^{2}$, respectively.
Namely, $m_{+}^{2}$ and $m_{-}^{2}$ are the squared masses of the
non-ghost and ghost modes, respectively. This identification can be
easily verified by considering the trivial case with $U_{,F\xi}=0$.
The condition for the model to be applicable to an experiment on a
scale $L_{{\rm exp}}$ will be $|m_{-}^{2}|\gg M_{{\rm cutoff}}^{2},$
with $M_{{\rm cutoff}}^{2}>L_{{\rm exp}}^{-2}$. If we do not see
any significant deviation from general relativity on this scale, a
condition $|m_{+}^{2}|> L_{{\rm exp}}^{-2}$ 
has to be imposed as well. These conditions are less intuitive. If
we are allowed to crudely identify $M_{{\rm cutoff}}^{2}$ with $L_{{\rm exp}}^{-2}$,
the above conditions are simplified to 
\begin{equation}
\left\vert 9\tilde{{\cal G}}\det(f_{,ij})\right\vert =\left\vert \frac{1}{m_{+}^{2}m_{-}^{2}}\right\vert \ll L_{{\rm exp}}^{4},\qquad\left\vert 3(f_{,RR}-\tilde{{\cal G}}f_{,{\cal GG}})\right\vert =\left\vert \frac{1}{m_{+}^{2}}+\frac{1}{m_{-}^{2}}\right\vert \ll L_{{\rm exp}}^{2}~,\label{eq:conditions}
\end{equation}
 where we used the fact that $m_{+}^{2}$ and $m_{-}^{2}$ are solutions
of Eq.~(\ref{eq:discrM2}).

Next we consider the case with $4\tilde{{\cal G}}U_{,F\xi}^{2}>(U_{,\xi\xi}+\tilde{{\cal G}}U_{,FF})^{2}$
(or equivalently $-4\tilde{{\cal G}}\det(f_{,ij})>(f_{,RR}-\tilde{{\cal G}}f_{,{\cal GG}})^{2}$),
in which we cannot simultaneously diagonalize both the kinetic and
mass matrices. In this case the eigenvalues $m^{2}$ obtained in Eq.~(\ref{msquare})
become complex. The complex nature of the solution means that those
modes are classically unstable. Since the eigenvalues for $m^{2}$
are complex conjugate with each other, a unique mass scale $\sqrt{|m^{2}|}$
must be much larger than the cutoff scale. Again, with the aid of
Eq.~(\ref{eq:discrM2}), this requirement leads to $-(\tilde{{\cal G}}\det f_{,ij})\lesssim L_{{\rm exp}}^{4}$.
This condition combined with $-4\tilde{{\cal G}}\det(f_{,ij})>(f_{,RR}-\tilde{{\cal G}}f_{,{\cal GG}})^{2}$
is identical to the conditions  of the previous case, given in (\ref{eq:conditions}).
Thus, we conclude that the conditions (\ref{eq:conditions}) are the
necessary and sufficient conditions for the model to be viable. When
we apply these constraints on the model to the solar system, we need
to plug in $R\approx(4\pi/3)\rho_{{\rm local}}\approx(4\pi/3)\times10^{-24}{\rm g/cm^{3}}\approx(10^{26}{\rm cm})^{-2}$
and ${\cal G}\approx48M_{\odot}^{2}/r^{6}\approx(5.7\times10^{16}{\rm cm})^{-4}(1{\rm AU}/r)^{6}$
as the background values. In the following we present two examples,
in which the expressions are simplified by assuming some hierarchy
among the components of the mass matrix $\tilde{M}_{ij}$.

The first case is the one in which the off-diagonal element $\tilde{M}_{12}$
is suppressed, i.e.\ $f_{,R\mathcal{G}}$ is negligible. If we set
$\tilde{M}_{12}=0$, the expressions for the mass eigenvalues simplify
to give
 \begin{equation}
m_{+}^{2}\approx\frac{U_{,FF}}{3}\approx\frac{1}{3f_{,RR}},\qquad m_{-}^{2}\approx-\frac{U_{,\xi\xi}}{3\tilde{{\cal G}}}\approx-\frac{1}{3\tilde{{\cal G}}f_{,{\cal GG}}}.
\end{equation}
The conditions that these masses are positive, $f_{,RR}>0$ and
$f_{,{\cal GG}}<0$, imply $\det(f_{,ij})<0$. Such a situation is
realized by considering the following type of
toy models of dark energy:
\begin{equation}
f(R,\mathcal{G})=R+{\cal F}_{R}(R)+{\cal F}_{\mathcal{G}}(\mathcal{G})\,,\label{de-model}
\end{equation}
where the functions ${\cal F}_{R}(R)$, and ${\cal F}_{\mathcal{G}}(\mathcal{G})$ have to be chosen such that $f_{,RR}>0$ and $f_{,GG}<0$. The correction ${\cal F}_{R}$ is of the kind discussed in
Refs.~\cite{Hu:2007nk,Starobinsky:2007hu,Tsujikawa:2007xu,Appleby:2007vb},
whereas ${\cal F}_{\mathcal{G}}$ is similar to the functions introduced in
Refs.~\cite{DeFelice:2008wz,Zhou:2009cy}. For concreteness, we specify
the functions as
 \begin{eqnarray*}
{\cal F}_{R}(R)=\mathcal{A}_{R}\mu^{2p+2}/(R^{p}+c_{R}\,\mu^{2p})~,\qquad{\cal F}_{\mathcal{G}}(\mathcal{G})=\mathcal{A}_{\mathcal{G}}\mu^{8n+2}/(\mathcal{G}^{2n}+c_{\mathcal{G}}\,\mu^{8n})~,
\end{eqnarray*}
where $\mu=O(H_{0})$ and $\mathcal{A}_{R},\mathcal{A}_{\mathcal{G}},c_{R},c_{\mathcal{G}}$
are constant parameters of $O(1)$. In these models, the corrections
to the cosmological evolution become important only at around the
present epoch\footnote{Here, ${\cal F}_{\mathcal{G}}$ is a good approximation to the models introduced
in \cite{DeFelice:2008wz} when ${\cal G}\gg H_{0}^{4}$, as it happens
in the vacuum Schwarzschild solution.%
}. When we consider local gravity in a dense region, the values of
$R$ and $\mathcal{G}$ are much larger than the cosmological backgrounds:
$R\gg H_{0}^{2}$ and $\mathcal{G}\gg H_{0}^{4}$. In this case, we
have $U_{,FF}=1/f_{,RR}=O(\mu^{-2p-2}\, R^{p+2})$ and $U_{,\xi\xi}=1/f_{,\mathcal{G}\mathcal{G}}=O(\mu^{-8n-2}\,\mathcal{G}^{2n+2})$,
and the mass squared for each mode is given by
 \begin{equation}
m_{\mbox{\scriptsize non-ghost}}^{2}=O\left(H_{0}^{2}{\left[\frac{\rho_{{\rm local}}}{\rho_{c}}\right]}^{p+2}\right),\quad m_{{\rm ghost}}^{2}=O\left(H_{0}^{2}\left[\frac{H_{0}^{-2}r_{s}}{r^{3}}\right]^{4n+2}\right),
\end{equation}
where $\rho_{c}\approx4\times10^{-30}$g/cm$^{3}$ is the critical
density of the universe. For the non-ghost mass, we have used the
Einstein equation, which is a good approximation in the local region
like the solar-system, to replace the Ricci scalar with the energy
density $\rho_{{\rm local}}$ of matter surrounding a star\footnote{Strictly speaking, one cannot take into account the effect of the
matter $\rho_{{\rm local}}$ in the present discussion since we have
assumed vacuum throughout the analysis. However, if the gravity is
dominated by the central star, which is true for the solar-system,
we expect that the gravity from the surrounding matter can be neglected
except for shifting the background values of $R$ in evaluating the
matrix elements of $U_{,ij}$ or $f_{,ij}$, and our vacuum results
can be used as the first approximation. %
}. If the system is exactly the vacuum, the non-ghost mode becomes
massless. Therefore, for the toy model of Eq.~(\ref{de-model}),
the non-ghost mode acquires mass through the chameleon mechanism.
For example, if we put the values for the solar system, $r_{s}\sim3{\rm km},~r\sim1{\rm AU},~\rho_{{\rm local}}\sim10^{-24}{\rm g/cm^{3}}$,
then we have $m_{\mbox{\scriptsize non-ghost}}^{2}\approx(10^{22.7-2.7p}{\rm cm})^{-2}$
and $m_{{\rm ghost}}^{2}\approx(10^{6-88.4n}{\rm cm})^{-2}$. One
can easily make the ghost mode sufficiently massive, while we need
to choose a relatively large power $p\geq4$ to make also the non-ghost
mode sufficiently massive, i.e.\ $m_{\mbox{\scriptsize non-ghost}}^{2}\gg(1{\rm AU})^{-2}$.

Interestingly, unlike the standard chameleon mechanism, we do not
need the matter to make the ghost mode very massive. However, the
background value of $\mathcal{G}$ is not always much higher than
the cosmological value. For example the value of the Gauss-Bonnet
term inside a star is typically negative (if evaluated, e.g.~on a
Schwarzschild interior solution) whereas it will be positive outside
the (neutron) star. Therefore there should be a point where $\mathcal{G}$
switches sign. At this point, the light ghost problem might arise.
However, since the region of the surface of the star where $\mathcal{G}=O(\mu^{4})$
might be very thin, the notion of the mass in the current treatment,
in which spatially homogeneous modes are assumed, becomes irrelevant.
Hence, it is not clear if actually there is a problem of light ghost
at this point. Another way around this problem consists of thinking
of a theory in which the background value of $R$ also contributes
to giving mass to the ghost when $\mathcal{G}$ is small. Some modification
of ${\cal F}_{\mathcal{G}}$ like ${\cal F}_{\mathcal{G}}\propto1/(\mathcal{G}^{2n}+cR^{q}+c_{\mathcal{G}}\mu^{8n})$
with a large power $q$ may work, but still some engineering for choosing the function will be needed.

The second case is when $|\tilde{{\cal G}}\det(f_{,ij})|\ll(f_{,RR}-\tilde{{\cal G}}f_{,\mathcal{GG}})^{2}$.
Basically, this is the case when the matrix $f_{,ij}$ is almost degenerate.
Once the condition $|\tilde{{\cal G}}\det(f_{,ij})|\ll(f_{,RR}-\tilde{{\cal G}}f_{,\mathcal{GG}})^{2}$
holds, the mass eigenvalues reduce to
 \begin{eqnarray*}
 &  & m_{\mbox{\scriptsize non-ghost}}^{2}\approx\frac{1}{3(f_{,RR}-\tilde{{\cal G}}f_{,\mathcal{GG}})}~,\quad m_{{\rm ghost}}^{2}\approx-\frac{f_{,RR}-\tilde{{\cal G}}f_{,\mathcal{GG}}}{3\tilde{{\cal G}}\det(f_{,ij})}~,~~~~~\mbox{for}~~~~|f_{,RR}|>|\tilde{{\cal G}}f_{,\mathcal{GG}}|,\\
 &  & m_{\mbox{\scriptsize non-ghost}}^{2}\approx-\frac{f_{,RR}-\tilde{{\cal G}}f_{,\mathcal{GG}}}{3\tilde{{\cal G}}\det(f_{,ij})}~,\quad m_{{\rm ghost}}^{2}\approx\frac{1}{3(f_{,RR}-\tilde{{\cal G}}f_{,\mathcal{GG}})}~,~~~~~\mbox{for}~~~~|\tilde{{\cal G}}f_{,\mathcal{GG}}|>|f_{,RR}|,\\
\end{eqnarray*}
 and the mass hierarchy $|m_{\mbox{\scriptsize ghost}}^{2}|\gg|m_{\mbox{\scriptsize non-ghost}}^{2}|$
is automatically guaranteed for $|f_{,RR}|>|\tilde{{\cal G}}f_{,\mathcal{GG}}|$.
In contrast, when $|\tilde{{\cal G}}f_{,\mathcal{GG}}|>|f_{,RR}|$,
the mass scale of the ghost mode is always lower than that of the
non-ghost mode. This limit includes a natural situation in which $f_{,RR}$,
$\mu^{2}f_{,R\mathcal{G}}$ and $\mu^{4}f_{,\mathcal{GG}}$ are the
same order with the typical energy scale $\mu$ satisfying $\mu^{4}\ll\mathcal{G}$.
In this case, we have $|\tilde{{\cal G}}f_{,\mathcal{GG}}|\gg|f_{,RR}|$
and therefore the mass scale of the ghost mode is lower than that
of the non-ghost mode. A more explicit expression for the masses are
$m_{\mbox{\scriptsize non-ghost}}^{2}\approx
f_{,\mathcal{GG}}/(3\det(f_{,ij}))$
and
$m_{\mbox{\scriptsize ghost}}^{2}\approx-1/(3\tilde{{\cal G}}f_{,\mathcal{GG}})$.
The expression for $m_{\mbox{\scriptsize non-ghost}}^{2}$ is slightly
different from the previous example in which we simply neglected $f_{,R\mathcal{G}}$.

\subsection{Case $\ell=0$.}

In the case with $j=0$ neither tensor
nor vector harmonics exist. Thus, 
the field $\alpha$ does not contribute
any longer to the action. Furthermore also the term quadratic in the
field $H_{1}$ disappears, together with the linear term in $H_{0}$.
In this case the variation with respect to $H_{1}$ 
gives a constraint which can be solved for $H_{2}$
in terms of $\delta F$, $\delta\xi$, $\delta F'$ and $\delta\xi'$.
At the same time also the contribution from $H'_{0}$ in action (\ref{eq:azioEv})
automatically cancels. Therefore we are left with an action in terms
of $\delta F$, $\delta\xi$ only, i.e.\ there are only two degrees
of freedom. 
In this case we find that the determinant
of the new kinetic matrix $A^{(l=0)}$ can be written as follows\begin{equation}
\det\left(A_{ij}^{(\ell=0)}\right)
 =-\frac{16\, r^{2}\,(2B-2-rB')^{2}}{A\, B\,(12B\xi'-2rF-4\xi'-r^{2}F')^{2}}<0\,.
\end{equation}
 Therefore also in this case there is one (and only one) ghost degree
of freedom. The speeds of propagation for both modes are equal to the
expressions given in Eq.~(\ref{eq:speed1}).

\subsection{Case $\ell=1$.}

In the case with $j^{2}=2$ the determinant in Eq.~(\ref{eq:detEv})
identically vanishes, as the tensor harmonics do not exist. This leads
to a reduction of the propagating degrees of freedom. In fact now
the matrix $A_{ij}^{(\ell=1)}$ possesses an eigenvector 
$\mathbf{N}_{2}$ 
with zero eigenvalue which satisfies 
$\mathbf{A}^{(\ell=1)}\,\mathbf{N}_{2}=\mathbf{0}$. 
This vector can be written as
\begin{equation}
\mathbf{N}_{2}
=\left(-\frac{3(4B\xi'-4\xi'-r^{2}F')}{r^{2}\,\xi'},\frac{\Gamma}{2AB\,(r\, F-4B\xi')},1\right)^{\! t},
\end{equation}
 where
\begin{eqnarray}
\Gamma & = & -4\, A\, B\, F\, r+2\, B\, F\, A'\,{r}^{2}+B\, A'\,{r}^{3}\, F'+4\, F\, r\, A\nonumber \\
 & - & 12\,{B}^{2}\xi'\, A'\, r+32\, A\,{B}^{2}\,\xi'-8\,\xi'\, B\, A\, r\, B'-32\, A\, B\,\xi'+4\, A'\, B\, r\,\xi'\,.
\end{eqnarray}
This suggests the field redefinition
\begin{eqnarray}
v_{1} & = & Q_{1}-\frac{3(4B\xi'-4\xi'-r^{2}F')}{r^{2}\,\xi'}\, X\,,\\
v_{2} & = & Q_{2}+\frac{\Gamma}{2AB\,(r\, F-4B\xi')}\, X\,,\\
v_{3} & = & X\:.
\end{eqnarray}
 In this way also the couplings between $X$ with $Q_{i}$, and $Q'_{i}$
vanish. The Lagrangian reduces to
\begin{equation}
\mathcal{L}=\tilde{A}_{ij}^{(\ell=1)}\,\dot{Q}_{i}\,\dot{Q}_{j}-\tilde{D}_{ij}^{(\ell=1)}\, Q'_{i}\, Q'_{j}-\tilde{M}_{ij}^{(\ell=1)}\, Q{}_{i}\, Q{}_{j}+C_{1}\, X^{2}+C_{i}X\, Q_{i}\,,
\end{equation}
(with $i,j=1,2$) so that by integrating out the field $X$, only
the mass term is affected. In this case we can find the ghost
condition and the speed of propagation. In particular
\begin{equation}
\det(\tilde{A}_{ij}^{(\ell=1)})=-\frac{144r^{2}\, A\, B\,(\xi')^{2}\,(2B-2-rB')^{2}}{\Gamma^{2}}\,,
\end{equation}
and therefore also in this case 
one of the two propagating degrees of freedom is a ghost. 
The speeds of propagation for the two modes
are equal and coincide with the expression given in Eq.~(\ref{eq:speed1}).

\section{Special cases}

\label{sp}

In the theories of gravity discussed so far, we have studied the general
function $f(R,\mathcal{G})$ for which
\begin{equation}
\Xi\equiv\frac{\partial^{2}f}{\partial R^{2}}\,\frac{\partial^{2}f}{\partial\mathcal{G}^{2}}-\left(\frac{\partial^{2}f}{\partial R\partial\mathcal{G}}\right)^{2}\neq0\,.
\end{equation}
 However there is a special class for which $\Xi$ identically vanishes,
e.g.\ the $f(R)$ theories. In general, $\Xi$ vanishes
when $F$ and $\xi$ are not independent, i.e.\ when $\xi=\xi(F)$.
When this happens, we have
\begin{equation}
\delta\xi=\frac{\xi'}{F'}\,\delta F\,,
\end{equation}
 and the independent scalar degrees of freedom reduce by one. In this
case one has to follow the same procedure as we have done for the
case $\Xi\neq0$. In particular now \begin{equation}
\delta F=v_{1}\,,
\end{equation}
 and $\alpha$ is substituted by $v_{0}$ according to
\[
j^{2}\, a_{4}\,\alpha=a_{4}\, v_{0}-a_{3}\, H_{2}-\frac{a_{4}\,(b_{2}\,\xi'+b_{3}\, F')}{4\, b_{1}\, F'}\, v'_{1}
\]
 by using the same relation (\ref{eq:alv0}). In order to remove higher
$r$-derivatives from the action one performs another field redefinition
as
\begin{eqnarray}
\psi & = & v_{1}\,,\\
v_{0} & = & (1+4j^{2})\, v_{2}\,.
\end{eqnarray}
 We have two no-ghost conditions which must be satisfied in order
to remove any ghost degree of freedom. In particular, the determinant
of the 2$\times$2 kinetic matrix $A_{ij}$ becomes
\begin{equation}
\det(A_{ij})=\frac{(j^{2}-2)\,(1+4j^{2})^{2}}{j^{2}}\,\frac{4A\, B\,[r^{2}\, F'-4(B-1)\,\xi']\,(F-2B'\,\xi'-4B\,\xi'')\,\Gamma_{2}}{(F')^{2}\,\Delta^{2}}\,,
\end{equation}
 where
\begin{eqnarray}
\Gamma_{2} & = & 3\,{r}^{2}FF'+2\,\xi'{r}^{2}F'B'-4\,\xi'F-16\,\xi'{B}^{2}\xi''+4\, B{r}^{2}F'\xi''\nonumber \\
 &  & +16\,\xi'B\xi''+24\,{\xi'}^{2}BB'-8\, r\xi'FB'+8\,{\xi'}^{2}B'-16\,\xi'BF'r+4\,\xi'BF\,,
\end{eqnarray}
 and $\Delta$ is defined in Eq.~(\ref{eq:Delta}). The other independent
condition, say $A_{11}\geq0$, is rather complicated, but for large
$j$ is given by
\begin{equation}
\lim_{j\to\infty}A_{11}=\frac{2\,(r\, F'-2B'\,\xi')\,[r^{2}\, F'-4(B-1)\,\xi']}{\sqrt{A\, B}\,(F')^{2}\,(r\, F-4B\,\xi')}\geq0\,,
\end{equation}
 so that in this case the ghost can be absent. As for the speeds of
the two modes, one coincides with the expression given in Eq.~(\ref{eq:speed1}),
whereas the other becomes
\begin{equation}
c^{2}=\frac{\Gamma_{3}}{A\,\Gamma_{2}}\,,
\end{equation}
 where
 \begin{eqnarray}
\Gamma_{3} & = & 24\,{\xi'}^{2}{B}^{2}A'+8\,{\xi'}^{2}BA'-4\,\xi'F\, A+2\,\xi'{r}^{2}F'A'B-8\,\xi'rA'F\, B-16\,\xi'F'rAB+4\,\xi'FAB+3\, F'{r}^{2}FA\:.
\end{eqnarray}

\section{Conclusion}

We have studied linear perturbations around the static spherically
symmetric spacetime for the modified gravity theories whose Lagrangian
consists of a general function $f(R,\mathcal{G})$ of both the Ricci
scalar $R$, and the Gauss-Bonnet scalar, $\mathcal{G}$%
\footnote{These theories are equivalent to the double-scalar-tensor
  theories defined by $\mathcal{L}=F\, R+\xi\,\mathcal{G}-U(F,\xi)$.%
}. For the odd-type modes, there are two degrees of freedom (one
dynamical variable). We derived the no gradient instability
condition. These conditions put constraints on the background
quantities.

For the even-type modes, the picture is more interesting.  We have
found that there are, in total, four degrees of freedom (corresponding
to two dynamical fields) for monopole and dipole perturbations
($\ell=0,1$), which do not exist in GR. Both for the monopole and
dipole perturbations, one of these two new scalar modes is always a
ghost. As for the higher multipole perturbations, there are, in total,
six degrees of freedom (corresponding to three dynamical fields), four
of which do not exist in GR. For Minkowski and (anti-)de Sitter
backgrounds, one out of these three fields is not independent from the
others.  Instead, for general backgrounds (including Schwarzschild), a
ghost always appears from the even-type perturbations with mass $m_-$
given in Eq.~(\ref{msquare}), whereas $m_+$ is the mass of the non-ghost
chameleon field. Finally the third dynamical field is massless describing the standard GR-mode.
The necessary and sufficient conditions that both the ghost and 
chameleon fields are sufficiently
massive, i.e.\ their inverse mass scale is smaller than 
$L_{{\rm exp}}$, the length scale of a determinate
experiment, reduces to~(\ref{eq:conditions}).
When these conditions hold, the ghost is massive enough to be treated
as a non-propagating mode in the effective gravitational theory. In
order to satisfy both these conditions, some hierarchy among the
$f_{,ij}$ coefficients is required, which means that some engineering
in choosing the function $f(R,{\cal G})$ is needed.

We also found that, in the high frequency limit, the radial
propagation speed of one field, among the three even-type modes,
coincides with the one for the odd-type mode, and the remaining two
even-type modes have a common propagation speed which is different
from the former. The classification of the modes are summarized in
Table I.

\begin{table}{}
\centering{}\begin{tabular}{|c|c|c|c|c|}
\hline 
 & odd-type G  & odd-type S  & even-type G  & even-type S \tabularnewline
\hline 
number of modes  & 0 ~(for $\ell=0,~1$)  & 0 ~(for $\ell=0,~1$)  & 2~(for $\ell=0,~1$)  & 1~(for $\ell=0,~1$) \tabularnewline
 & 1~(for $\ell\ge2$)  & 1~(for $\ell\ge2$)  & 3~(for $\ell\ge2$)  & 2~(for $\ell\ge2$) \tabularnewline
\hline 
ghost and  & $AF-2A'B\xi'>0$  & $AF-2A'B\xi'>0$  & ghost present  & constraints\tabularnewline
gradient conditions & $F-4B\xi''-2B'\xi'>0$  & $F-4B\xi''-2B'\xi'>0$  & massive if Eq.~(\ref{eq:conditions}) is verified & (see Sec.\ref{sp}) \tabularnewline
\hline 
\end{tabular}\caption{Classification of the modes for the general $f(R,\mathcal{G})$ theories.
G and S stand for the model that does or does not satisfy Eq.~(\ref{spcond}).}
\end{table}

If the theory satisfies the condition:
 \begin{equation}
\frac{\partial^{2}f}{\partial R^{2}}\frac{\partial^{2}f}{\partial G^{2}}-{\left(\frac{\partial^{2}f}{\partial R\partial G}\right)}^{2}=0,\label{spcond}
\end{equation}
then the number of modes for the even-type perturbations reduces
by one. In this case, the ghost is not necessarily present, i.e.~we
get non-trivial no-ghost condition which puts bound on the function
form of $f(R,\mathcal{G})$ and also the background spacetime. For
example, in $f(R)$ gravity theories, the no-ghost condition is 
satisfied once $\partial f/\partial R=F>0$, and we find that 
the ghost does not exist in this case as expected. 

\begin{acknowledgments}
  The work of A.\,D.\,F.\ was supported by the Grant-in-Aid for Scientific
  Research Fund of the JSPS No.~10271. A.\,D.\,F.\ is also grateful to
  Yukawa Institute for warm hospitality during his stay in Yukawa
  Institute, especially during the workshop YITP-W-10-10.  T.\,S.\ is
  supported by a Grant-in-Aid for JSPS Fellows No.\ 1008477.
  T.\,T.\ is supported by JSPS Grant-in-Aid for Scientific Research (A)
  No.\ 21244033, the Global COE Program ``Next Generation of Physics,
  Spun from Universality and Emergence,'' and the Grant-in-Aid for
  Scientific Research on Innovative Areas No.\ 21111006 from the MEXT.
\end{acknowledgments}

\newpage

\appendix

\section{Coefficients in the action}

We define here the different coefficients introduced in order to define
the action. By calling $\Theta\equiv r^{2}\sqrt{A/B}$, we have\begin{eqnarray}
a_{1} & = & -a_{4}\, b_{2}/(4b_{1})\,,\\
a_{2} & = & -a_{4}\, b_{3}/(4b_{1})\,,\\
a_{3} & = & -a_{4}\, b_{4}/(4b_{1})\,,\\
a_{4} & = & {\Theta B\left(r\, F-4\, B\xi'\right)/r^{3}}\,,\\
a_{5} & = & b_{6}\, a_{6}/b_{7}\,,\\
a_{6} & = & -{\Theta B\, A'/(2A)}\,,\\
a_{7} & = & a_{4}\, c_{5}/b_{2}\,,\\
a_{8} & = & -{\Theta B\left(4\, B\xi'-F\, r\right)\left(A'\, r-2\, A\right)/(2Ar^{4})}\,,\\
a_{9} & = & -{2\Theta B'/r^{3}}\,,\\
a_{10} & = & c_{9}\,,\\
b_{1} & = & {\Theta B\,\left(F\, r-4\, B\xi'\right)/(2Ar^{3})}\,,\\
b_{2} & = & {8B\,\Theta\left(B-1\right)/(Ar^{2})}\,,\\
b_{3} & = & -{\frac{2\Theta B}{A}}\,,\\
b_{4} & = & -{\Theta B\left(-F'\, r^{2}-4\,\xi'-2\, F\, r+12\, B\,\xi'\right)/(Ar^{2})}\,,\\
b_{5} & = & -2\, b_{1},\\
b_{6} & = & b_{2}\, b_{7}/b_{3}\,,\\
b_{7} & = & -4b_{1}\, a_{6}/a_{4}\,,\\
c_{1} & = & -{B\Theta\left(24\, B\,\xi'A'-4\,\xi'A'-r^{2}F'\, A'-2\, F\, r\, A'-2\, A\, F-4\, A\, r\, F'\right)/(4Ar^{2})}\,,\\
c_{2} & = & {2\Theta B\, A'\left(3B-1\right)/(Ar^{2})}\,,\\
c_{3} & = & -{B\Theta\left(rA'+4\, A\right)/(2Ar)}\\
c_{4} & = & {B\Theta\left(12\, B\xi'A'-2\, A\, r\, F'-2\, A\, F-F\, r\, A'\right)/(2Ar^{3})}\,,\\
c_{5} & = & {4\Theta(B-1)/(Ar^{2})}\,,\\
c_{6} & = & d_{3}\, c_{5}/b_{2}\,,\\
c_{7} & = & -{\Theta(B{A'}^{2}-2\, A\, B\, A''-{B}^{2}{A'}^{2}+3\, A\, B\, B'A'+2\,{B}^{2}A\, A''-AB'A')/(A^{2}r^{2})}\,,\\
c_{8} & = & b_{3}\, c_{5}/b_{2}\,,\\
c_{9} & = & \Theta/r^{2}\,,\\
c_{10} & = & {\Theta(4\,{A}^{2}B'-B{A'}^{2}r+2\, A\, B\, A''r+AB'A'r)/(4rA^{2})}\,,\\
d_{1} & = & b_{1}\,,\\
d_{2} & = & -{\Theta B\left(2\, B\,\xi'\, A'-A\, F\right)/(Ar^{4})}\,,\\
d_{3} & = & -{4\Theta{B}^{2}A'/(Ar^{3})}\\
d_{4} & = & c_{9}\, b_{2}/c_{5}\,,\\
d_{5} & = & {4\Theta{B}^{2}A'/(Ar^{4})}\,,\\
d_{6} & = & -{2B\Theta/r^{3}}\,,\\
e_{1} & = & -\Theta{\it U}_{,FF}/2\,,\\
e_{2} & = & -\Theta U_{,F\xi}\,,\\
f_{1} & = & -\Theta{\it U}_{,\xi\xi}/2\,.
\end{eqnarray}
 The elements of the kinetic matrix relevant for our discussion are\begin{eqnarray}
K_{22} & = & \frac{1+4j^{2}}{j^{2}\,(4c_{5}a_{4}b_{1}j^{2}+a_{8}b_{2}b_{4})^{2}}\left[2\,{b_{{2}}^{2}}{b_{{4}}^{2}}\left(b_{{1}}a'_{{8}}a_{{4}}+b_{{1}}a_{{8}}a'_{{4}}-2\, b_{{1}}{a_{{8}}^{2}}-b'_{{1}}a_{{8}}a_{{4}}\right)\right.\nonumber \\
 & - & \left.8\, j^{2}{b_{{1}}^{2}}a_{{4}}\left(a_{{4}}c_{{5}}b'_{{2}}b_{{4}}+a_{{4}}c_{{5}}b_{{2}}b'_{{4}}-a_{{4}}b_{{2}}c'_{{5}}b_{{4}}-2\, c_{{5}}b_{{2}}b_{{4}}a'_{{4}}+2\, c_{{5}}a_{{8}}b_{{2}}b_{{4}}\right)\right]\\
K_{33} & = & \frac{(1+4j^{2})\,(b_{2}b'_{3}-b_{3}b'_{2})^{2}}{8j^{2}\, b_{1}^{2}\, b_{3}^{2}\, r\,(4c_{5}a_{4}b_{1}j^{2}+a_{8}b_{2}b_{4})^{2}}\left[-8\, j^{4}{b_{{1}}^{3}}a_{{4}}c_{{5}}\left(4\, a_{{4}}r\, c_{{5}}-{c_{{9}}^{2}}b_{{4}}b_{{2}}\right)\right.\nonumber \\
 & - & 2\, j^{2}{b_{{1}}^{2}}\left(2\,{a_{{4}}^{2}}c_{{5}}b_{{4}}rb'_{{2}}-4\, a_{{4}}c_{{5}}b_{{2}}b_{{4}}a'_{{4}}r+2\,{a_{{4}}^{2}}c_{{5}}b_{{2}}b'_{{4}}r-{b_{{2}}}^{2}{b_{{4}}}^{2}a_{{8}}{c_{{9}}}^{2}+8\, a_{{4}}c_{{5}}a_{{8}}b_{{2}}b_{{4}}r-2\,{a_{{4}}^{2}}b_{{2}}b_{{4}}c'_{{5}}r\right)\nonumber \\
 & + & \left.{b_{{2}}^{2}}{b_{{4}}^{2}}r\left(b_{{1}}a_{{8}}a'_{{4}}-2\, b_{{1}}{a_{{8}}^{2}}+b_{{1}}a_{{4}}a'_{{8}}-a_{{4}}a_{{8}}b'_{{1}}\right)\right],
\end{eqnarray}
 where\[
b_{2}b'_{3}-b_{3}b'_{2}\propto r\, B'-2B+2\,,
\]
 which vanishes on (anti-)de Sitter and on Minkowski.

For backgrounds close to the Schwarzschild one, the matrix $K_{ij}$
reduces to\begin{eqnarray*}
K_{11} & = & {\frac{\left(2\,{r}^{2}{j}^{6}+\left(10\, r_{{s}}-7r\right)r{j}^{4}+\left(14\,{r_{{s}}}^{2}-20\, rr_{{s}}+6\,{r}^{2}\right){j}^{2}-{r_{{s}}}^{2}\right){r}^{3}}{{j}^{2}\,\left(r{j}^{2}-2\, r+3\, r_{{s}}\right)^{2}\left(r-r_{{s}}\right)}}\,,\\
K_{12} & = & -{\frac{{r}^{2}\sqrt{1+4\,{j}^{2}}\left({j}^{2}-2\right)\left(r{j}^{2}+r_{{s}}\right)}{{j}^{2}\left(r{j}^{2}-2\, r+3\, r_{{s}}\right)^{2}}}\,,\\
K_{13} & = & -{\frac{6\left({r}^{2}{j}^{6}+2\left(2\, r_{{s}}-r\right)r{j}^{4}+\left(7\,{r_{{s}}}^{2}-6\, rr_{{s}}\right){j}^{2}-4\, rr_{{s}}+4\,{r_{{s}}}^{2}\right)r_{{s}}\sqrt{1+4\,{j}^{2}}}{{j}^{2}\,\left(r{j}^{2}-2\, r+3\, r_{{s}}\right)^{2}\left(r-r_{{s}}\right)}}\,,\\
K_{22} & = & {\frac{2\left(r-r_{{s}}\right)r\left(1+4\,{j}^{2}\right)\left({j}^{2}-2\right)}{{j}^{2}\left(r{j}^{2}-2\, r+3\, r_{{s}}\right)^{2}}}\,,\\
K_{23} & = & \frac{12r_{s}}{r^{2}}\, K_{22}\,,\\
K_{33} & = & \frac{12r_{s}}{r^{2}}\, K_{23}\,.
\end{eqnarray*}

\bibliographystyle{apsrev}
\bibliography{fRGs}

\end{document}